\documentclass[11pt]{article}

\usepackage[top = 30 mm, bottom = 30mm, left = 20mm, right=20mm]{geometry}
\usepackage{amsmath, amssymb}
\usepackage[english]{babel}
\usepackage[utf8]{inputenc}
\usepackage{latexsym}
\usepackage{graphicx}
\usepackage{lmodern}
\usepackage{jheppub}
\usepackage{slashed}
\usepackage{charter}
\usepackage{xcolor}
\usepackage{color}
\usepackage{float}
\usepackage{bm}
\usepackage{braket}

\numberwithin{equation}{section}

\newcommand{\be}{\begin{equation}}
	\newcommand{\ee}{\end{equation}}
\newcommand{\bear}{\begin{array}}
	\newcommand{\eear}{\end{array}}

\title{More on massive gravitino scattering amplitudes and\\the unitarity cutoff of the new Fayet-Iliopoulos terms}
\author[a,b]{Ignatios Antoniadis}
\author[a]{Jules Cunat}
\author[a]{Anthony Guillen}

\affiliation[a]{Laboratoire de Physique Th\'eorique et Hautes Energies - LPTHE\\
	Sorbonne Universit\'e, CNRS, 4 Place Jussieu, 75005 Paris, France}

\affiliation[b]{Department of Physics, Harvard University, Cambridge, MA 02138, USA}

\emailAdd{antoniad@lpthe.jussieu.fr}
\emailAdd{jcunat@lpthe.jussieu.fr}
\emailAdd{aguillen@lpthe.jussieu.fr}

\date{} 

\abstract{We extend the $2\rightarrow2$ gravitino scattering amplitude computed in \cite{Antoniadis:2022jjy} to an arbitrary $\mathcal{N}=1$ supergravity model of one chiral and one vector multiplet, in a Minkowski background with supersymmetry breaking driven by both $F$- and $D$-terms. We find that the cancellation of the leading term in $\mathcal{O}(\kappa^2 E^4/|m_{3/2}|^2)$, that would lead to a breakdown of perturbative unitarity at a scale $\Lambda\sim M_\mathrm{SUSY}$, is a consequence of the vanishing of the scalar potential at its minimum, which is implied by the flat background. We then analyse the inclusion of the new Fayet-Iliopoulos (FI) terms. We find that, since they modify the scalar potential without contributing to the amplitudes, they generically lead to uncanceled leading terms in the latter and a perturbative cutoff at the supersymmetry breaking scale, except for particular cases where the new FI term does not modify the potential at its minimum and the cutoff is pushed up to the Planck scale.}

\begin{document}
	
	\maketitle\newpage\vfill
	
	\section{Introduction}
	
	\vspace{1\baselineskip}
	
	In a recent paper \cite{Antoniadis:2022jjy}, we started an investigation of the massive gravitino $2\rightarrow2$ scattering amplitudes. We computed them in the Polonyi model \cite{Polonyi:1977pj} and found that, as for the standard Higgs mechanism, there is a cancellation of the leading terms at high energy (in $\mathcal{O}(\kappa^2 E^4/|m_{3/2}|^2)$, where $E$ is the energy of the gravitinos and $m_{3/2}$ their mass) between gravitational and scalar channels, allowing the perturbative unitarity cutoff to lie at the Planck scale. We then added an abelian vector multiplet without charging the Polonyi field, and considered the effect of the new Fayet-Iliopoulos (FI) term that was introduced in \cite{Cribiori:2017laj, Antoniadis:2018cpq, Antoniadis:2018oeh, Cribiori:2018dlc}, as a way to emulate the standard supergravity FI term \cite{Fayet:1974jb, Freedman:1976uk} without gauged R-symmetry and its issues \cite{Komargodski:2009pc,Dienes:2009td}. We found that in this case, the aforementioned cancellation cannot happen because this new FI term does not contribute to the amplitudes, leading to a cutoff at the supersymmetry (SUSY) breaking scale, associated with the vector $D$-auxiliary component expectation value. This result is in agreement with the analysis of \cite{Jang:2022sql,Jang:2020cbe} using different arguments.
	
	\vspace{1\baselineskip}
	
	The present paper has two goals. The first is to extend the amplitude computation previously made in the Polonyi model to an arbitrary model of one vector and one charged chiral multiplet, gauging in general the R-symmetry, so that supersymmetry can also be broken by a $D$-term, generalising the Polonyi model which has only $F$-term SUSY breaking. This is done in section 2. We obtain that, for amplitudes computed around some minimum of the potential in Minkowski gravitational background, the condition for cancellation is precisely the vanishing of the potential at its minimum, which is implied by the flat background. In section 3, we analyse the inclusion of new Fayet-Iliopoulos terms in this setup. We consider two cases that we call the \emph{original new FI term} and the \emph{K\"ahler invariant new FI term}. The \emph{original new FI term} is the one introduced in \cite{Cribiori:2017laj, Cribiori:2018dlc} with matter couplings that are not K\"ahler invariant \cite{Antoniadis:2018cpq}, while the \emph{K\"ahler invariant new FI term} consists of a modification proposed in \cite{Antoniadis:2018oeh}. Both terms have been used for cosmological applications, for instance in \cite{Antoniadis:2018cpq, Antoniadis:2019nwz, Jang:2021fce, Jang:2022sql}. Also, to avoid confusion, we call \emph{standard FI term} the one from standard supersymmetry \cite{Freedman:1976uk} which implies a gauged R-symmetry. As in \cite{Antoniadis:2022jjy}, we find that since the new FI terms modify the scalar potential without contributing to the amplitudes, they generically lead to uncanceled leading terms $\mathcal{O}(\kappa^2 E^4/|m_{3/2}|^2)$ in the latter and a perturbative cutoff at the SUSY breaking scale. In concrete examples, we can find particular values of the parameters for which the contribution of the new FI terms to the potential vanishes at the minimum, restoring the cancellation in the amplitudes and a cutoff at the Planck scale. In general, this phenomenon requires an extra tuning of parameters, besides the one of vacuum energy, unless we consider the new FI term to be field dependent, which leads to a rather trivial result.

	\vfill\newpage\vfill
	
	\section{Gravitino scattering with chiral and vector multiplets}
	
	\vfill
	
	In \cite{Antoniadis:2022jjy}, we computed the massive gravitino scattering amplitudes in the Polonyi model, which includes a chiral multiplet with canonical K\"ahler potential and a constant+linear superpotential. There are two channels contributing to these amplitudes at tree level, with propagation of the graviton or the scalar comprised in the chiral multiplet. We found that, in a similar way as for the standard Higgs mechanism, there is a cancellation of the leading terms at high energy (in $\mathcal{O}(\kappa^2 E^4/|m_{3/2}|^2)$) between these channels, pushing the perturbative unitarity cutoff up to the Planck scale. The computation was done in Minkowski vacuum with supersymmetry broken by a vacuum expectation value (VEV) of the chiral multiplet $F$-auxiliary component.
	
	\vfill
	
	The goal of this section is to generalise this result for one vector and one charged chiral multiplet, with arbitrary K\"ahler potential $\mathcal{K}(z, \bar z)$, superpotential $W(z)$, and gauge kinetic function $f(z)$. We assume that spontaneous supersymmetry breaking takes place and use the unitary gauge, where the mixing between the gravitino and the Goldstino combination of the two spin-1/2 fermions is set to zero. This time, there are three channels contributing to the $2\rightarrow2$ gravitino scattering at tree level, with propagation of the graviton, the scalar, and the gauge vector, in their respective multiplets. The contribution of the gravitational channel is not modified with respect to \cite{Antoniadis:2022jjy} in this setup; we will recall the result later. The contributions to compute are the ones from the scalar and vector channels. For this, the relevant terms in the $\mathcal{N}=1$ supergravity Lagrangian are \cite{Freedman:2012zz}
	\begin{eqnarray}\label{eq:lagrangian_N=1}
		e^{-1}\mathcal{L}\supset 
		&-&\frac{1}{2}\bar\psi_\mu\gamma^{\mu\rho\sigma}\left(\partial_\rho - \frac{3}{2}i\mathcal{A}_\rho\gamma_*\right)\psi_\sigma
		-\partial\bar\partial\mathcal{K}\hat{\partial}^\mu z\hat{\partial}_\mu\bar z
		- \frac{1}{4} \mathrm{Re}(f) F^{\mu\nu}F_{\mu\nu}\nonumber\\
		&+& \frac{1}{2}(e^{\mathcal{K}/2}W\bar\psi_\mu P_R\gamma^{\mu\nu}\psi_\nu + \mathrm{h.c.}).
	\end{eqnarray}
	
	Here, we did set the gravitational coupling $\kappa=1$, and use standard notations for the different fields involved. Partial derivatives without index $\partial$ ($\bar\partial$) stand for differentiation with respect to the scalar field $z$ ($\bar z$), $A_\mu$ is the gauge potential with field strength $F_{\mu\nu}$, $\psi_\mu$ is the gravitino, while $\gamma^{\mu\nu}$ ($\gamma^{\mu\rho\sigma}$) denotes the totally antisymmetric product of two (three) Dirac gamma-matrices. Also, $\gamma_* = i\gamma_0\gamma_1\gamma_2\gamma_3$ is used to define the chiral projector $P_R = 1/2(1 - \gamma_*)$. The covariant derivative of the scalar is given by $ \hat{\partial}_{\mu}z=\partial_{\mu}z-A_{\mu}k$ where $k(z)$ is the Killing vector involved in the transformation of the scalar $z$ (with only one vector multiplet, the symmetry is $U(1)$ and $k(z) = -iqz$, with $q$ the charge of the scalar). Finally, $\mathcal{A}_\mu$ is the K\"ahler connection given by
	\begin{eqnarray}\label{eq:connection_Kahler}
		\mathcal{A}_\mu &=& \frac{i}{6}\left(\partial_\mu z\partial\mathcal{K} - \partial_\mu\bar z\bar\partial\mathcal{K}\right) - \frac{1}{3}A_\mu\mathcal{P}=\frac{i}{6}\left(\hat{\partial}_\mu z\partial\mathcal{K} - \hat{\partial}_\mu\bar z\bar\partial\mathcal{K}+A_{\mu}(r-\bar{r})\right).
	\end{eqnarray}
	
	\vfill\newpage\vfill
	
	The moment map $\mathcal{P}(z, \bar z)$ appearing there is a real function such that $\bar\partial\mathcal{P}(z, \bar z)=ik(z)\partial\bar\partial\mathcal{K}$ and $r(z)$ is defined from the gauge transformation of the K\"ahler potential, as $(k(z)\partial+\bar k(\bar z)\bar\partial)\mathcal{K}(z, \bar z)=r(z)+\bar r(\bar z)$. The (gauge invariant) real moment map can then be written 
	\begin{equation}\label{eq:definition_P}
		\mathcal{P}(z, \bar z) = i(k(z)\partial\mathcal{K}(z, \bar z) - r(z)) = -i(\bar k(\bar z)\bar \partial\mathcal{K}(z, \bar z) - \bar r ( \bar z)).
	\end{equation}

	Before giving the result for the amplitudes, let us remind some standard facts on the Fayet-Iliopoulos term, R-symmetry and K\"ahler transformations. As one can see, the definition of $r(z)$ does not constrain its imaginary part, so one is free to add a constant to it, which translates as a real constant in $\mathcal{P}(z, \bar z)$. This arbitrary constant is the so-called Fayet-Iliopoulos constant, and it can be added to the moment map of any $U(1)$. If we start from $\mathrm{Im}(r(z)) = 0$, adding this constant for a given $U(1)$ has for consequence that it becomes an R-symmetry. It can be seen for instance in the gauge transformation of the gravitino, which is proportional to $\bar r(\bar z) - r(z)$
	\begin{equation}\label{eq:transformation_gravitino}
		\delta\psi_\mu = \frac{1}{4}(\bar r(\bar z) - r(z))\psi_\mu\theta.
	\end{equation}
	
	One can check that the first term of \eqref{eq:lagrangian_N=1} is gauge invariant since the gauge transformation of $\mathcal{A}_\mu$ given in \eqref{eq:connection_Kahler} is
	\begin{equation}
		\delta \mathcal{A}_\mu = -\frac{i}{6}\partial_\mu\left((\bar r(\bar z) -r(z))\theta\right).
	\end{equation}
	
	This transformation of the gravitino can be traced back to the transformation of the compensator multiplet $S_0$ in the superconformal construction of $\mathcal{N}=1$ supergravity, before gauge fixing the chiral T-symmetry of the superconformal algebra using the condition $S_0 = \bar S_0$.
	
	\vfill
	
	In other words, $r(z)$ can be regarded as the Killing vector involved in the transformation of the compensator $S_0$. As soon as Im$(r(z)) \neq 0$, the field $S_0$ transforms under the $U(1)$ and this defines an R-symmetry (which is equivalent to a phase transformation of the fermionic coordinates in the superfield formalism). Moreover, $r(z)$ is also involved in the gauge transformation of the superpotential
	\begin{equation}\label{eq:gauge-transformation_superpotential}
		\delta W(z) = k(z)\partial W(z)\theta = -r(z)W(z)\theta,
	\end{equation}
	
	which constrains the form of the superpotential and the charges of the scalars. For instance, if $r(z) = i\xi_s$, the superpotential must transform as $W(z)\rightarrow \exp(-i\xi_s\theta)W(z)$ and the only possibility is $W(z) = z^b$ with the charge $qb = \xi_s$. In this way, R-symmetry can be seen as a symmetry under which $W(z)$ transforms, and this can be taken as an equivalent definition. On the other hand, the real part Re$(r(z))$ is associated to a local scale transformation before it is fixed by a condition on $|S_0|$, usually chosen to have canonically normalised gravity kinetic terms in the Einstein frame.
	
	\vfill\newpage\vfill
	
	Now, note that the Killing vector $r(z)$ is not invariant under K\"ahler transformation. Indeed, since under such transformations $\mathcal{K}(z, \bar z) \rightarrow \mathcal{K}(z, \bar z) + J(z) + \bar J(\bar z)$ we can easily obtain 
	\begin{equation}\label{eq:Kahler_Killing}
		r(z) \rightarrow r(z) + k(z)\partial J(z),
	\end{equation}
	
	implying that it is always possible to choose $\partial J(z) = -r(z)/k(z)$ such that $r(z) \rightarrow 0$ and the $U(1)$ becomes an ordinary non R-symmetry. From \eqref{eq:gauge-transformation_superpotential}, we see that $-r(z)/k(z) = \partial\log W(z)$, so $J(z) = \log W(z) + \mathrm{constant}$, and the associated K\"ahler transformation of the superpotential $W(z) \rightarrow e^{-J}W(z)$ makes it constant. In particular it does not transform under the $U(1)$ anymore. In short, R-symmetry is not a K\"ahler frame independent concept; we can always go to a K\"ahler frame where the superpotential is constant and the gauge symmetry is not an R-symmetry. Note however that the moment map $\mathcal{P}(z, \bar z)$ is invariant under K\"ahler transformations; so if the standard FI constant is added in some Kahler frame it is present in any other K\"ahler frame, even in the one where the $U(1)$ is not a R-symmetry. For instance, in some frames it can get incorporated into the K\"ahler potential.
	
	\vfill
	
	Let us now come back to our amplitudes. To start, let us suppose that the scalar $z$ picks a nonvanishing vev with $|z_0| = v$ at the minimum of the potential, and parameterise it as
	\begin{equation}\label{eq:minim_potential}
		z(x) = (v+\eta(x))e^{i\phi(x)}, \quad \mathrm{where}\quad \partial V(z_0, \bar z_0)=\bar\partial V(z_0, \bar z_0) = 0\quad \mathrm{and}\quad \left<\eta(x)\right>=0.
	\end{equation}
	
	Since $z$ is charged, this vev breaks the $U(1)$, and in the unitary gauge (we use the term "unitary gauge" for both SUSY and the $U(1)$, but the distinction should be clear with context), the phase $\phi(x)$ can be reabsorbed in the gauge potential $A_\mu$ that becomes massive, with its mass term contained in the kinetic term of $z$ in \eqref{eq:lagrangian_N=1}. In this gauge, we simply have $z(x) = v + \eta(x)$, which is real. The mass of $A_\mu$ is given by
	\begin{equation}
		M_A^2 = 2(qv)^2 \frac{\partial\bar\partial\mathcal{K}_0}{\mathrm{Re}(f_0)} = \frac{\partial\bar\partial\mathcal{K}_0W_0\bar W_0}{\nabla W_0\bar\nabla \bar W_0}\frac{2\mathcal{P}_0^2}{\mathrm{Re}(f_0)},
	\end{equation}
	
	where $\nabla$ is the K\"ahler covariant derivative $\nabla W = \partial W + (\partial\mathcal{K})W$ and the subscript $0$ means evaluated at $(z_0, \bar z_0)$. In this expression, we redefined the gauge potential to be canonically normalised, which at lowest order amounts to $A^\mu \rightarrow A^\mu/(\mathrm{Re}(f_0))^{1/2}$; this is where the $\mathrm{Re}(f_0)$ comes from. In the second equality, we used the definition of $\mathcal{P}$ in \eqref{eq:definition_P}, along with $k = -iqz \sim -iqv$ at the minimum, and $-r/k = \partial\log W$. With the unitary gauge, the propagator of the vector becomes 
	\begin{equation}
		P_A^{\alpha\beta}(k) = -\frac{i(\eta^{\alpha\beta}+k^\alpha k^\beta/M_A^2)}{k^2+M_A^2}.
	\end{equation}
	
	\vfill\newpage\vfill
	
	Now, to compute the amplitudes we need to expand $e^{\mathcal{K}/2}W$ to first order in $\eta$, as needed to obtain the 3-point gravitino interactions $\eta\psi\psi$ and $A\psi\psi$. We then canonically normalise $\eta$ just like $A$, by the redefinition $\eta \rightarrow \eta/(2\partial\bar\partial\mathcal{K}_0)^{1/2}$. We obtain the following interactions
	\begin{equation}\label{eq:lagrangian_interaction_scalar}
		\mathcal{L}_{\eta\psi\psi}^{(1)} = -\frac{1}{8\sqrt{2}}\frac{\partial\mathcal{K}_0-\bar\partial\mathcal{K}_0}{\sqrt{\partial\bar\partial\mathcal{K}}_0}(\partial_\rho \eta) \bar\psi_\mu\gamma^{\mu\rho\sigma}\gamma_*\psi_\sigma,
	\end{equation}
	
	and
	\begin{eqnarray}
		\mathcal{L}_{\eta\psi\psi}^{(2)} =
		&+&\frac{e^{\mathcal{K}_0/2}}{8\sqrt{2}}\frac{(\partial\mathcal{K}_0+\bar\partial\mathcal{K}_0)(W_0+\bar W_0)+2(W_0+\bar W_0)}{\sqrt{\partial\bar\partial\mathcal{K}_0}}\eta\bar\psi_\mu\gamma^{\mu\nu}\psi_\nu\nonumber\\
		&-&\frac{e^{\mathcal{K}_0/2}}{8\sqrt{2}}\frac{(\partial\mathcal{K}_0+\bar\partial\mathcal{K}_0)(W_0-\bar W_0)+2(W_0-\bar W_0)}{\sqrt{\partial\bar\partial\mathcal{K}_0}}\eta\bar\psi_\mu\gamma^{\mu\nu}\gamma_*\psi_\nu.
	\end{eqnarray}
	
	We also get
	\begin{equation}\label{eq:lagrangian_interaction_vector}
		\mathcal{L}_{A\psi\psi} = -\frac{i}{4}\frac{\mathcal{P}_0}{\sqrt{\mathrm{Re}(f_0)}}A_\rho\bar\psi_\mu\gamma^{\mu\rho\sigma}\gamma_*\psi_\sigma.
	\end{equation}
	
	Note that even if the gravitino is not charged under the $U(1)$ (i.e. $\mathrm{Im}(r(z)) = 0$ in \eqref{eq:transformation_gravitino}), the latter can have an interaction $A\psi\psi$ with the gauge vector through $\mathcal{P}_0$ when the scalar has a vacuum expectation value.
	
	\vfill 
	
	These interactions can then straightforwardly be turned into vertices and use them to compute the $2\rightarrow2$ gravitino scattering amplitude. The details of this computation are as in \cite{Antoniadis:2022jjy}. An important point is that it is done in a Minkowski background; in other words $V(z_0, \bar z_0) = 0$.
	
	\vfill
	
	We end up with the following contributions from the scalar and vector channels, for instance to the $(+,+,-,-)$ amplitude, with two external helicities $+1/2$ and two helicities $-1/2$ (we consider amplitudes with external helicities $\pm1/2$ because they are the ones which diverge the most at high energy, as explained in \cite{Antoniadis:2022jjy})
	\begin{equation}
		\mathcal{M}^{+, +, -, -}_\mathrm{scalar} = -\frac{8\kappa^2 E^4}{9|m_{3/2}|^2}\frac{\nabla W_0 \bar \nabla \bar W_0}{W_0\bar W_0\partial\bar\partial\mathcal{K}_0} + \mathcal{O}(\kappa^2 E^2),
	\end{equation}
	
	and
	\begin{equation}
		\mathcal{M}^{+, +, -, -}_\mathrm{vector} = -\frac{8\kappa^2E^4}{9|m_{3/2}|^2}\left(\frac{2}{M_A^2}+\frac{1}{|m_{3/2}|^2}\right)\frac{\mathcal{P}_0^2}{\mathrm{Re}(f_0)} + \mathcal{O}(\kappa^2 E^2),
	\end{equation}
	
	where we restored the $\kappa$-dependence. We also recall the result for the gravitational channel~\cite{Antoniadis:2022jjy}
	\begin{equation}
		\mathcal{M}^{+, +, -, -}_\mathrm{grav} = \frac{16\kappa^2E^4}{3|m_{3/2}|^2} + \mathcal{O}(\kappa^2 E^2).
	\end{equation}
	
	In all these expressions, $E$ is the energy of the gravitinos and $m_{3/2} = e^{\mathcal{K}_0/2}W_0$ their mass. 
	
	\vfill\newpage\vfill
	
	These amplitudes exhibit a divergent behaviour at high energy. If we look at the different channels individually, we can expect perturbative unitarity breakdown when $\mathcal{M}\sim 1$, at the supersymmetry breaking scale $\Lambda \sim (m_{3/2}/\kappa)^{1/2}\sim M_{\mathrm{SUSY}}$. However, when summing them
	\begin{equation}\label{eq:divcancel}
		\mathcal{M}_{\mathrm{total}}^{+, +, -, -} = -\frac{16\kappa^2E^4}{9|m_{3/2}|^4}\left(e^{\mathcal{K}_0}\left(\frac{\nabla W_0\bar \nabla \bar W_0}{\partial\bar\partial\mathcal{K}_0}-3W_0\bar W_0\right) + \frac{\mathcal{P}_0^2}{2\mathrm{Re}(f_0)}\right) + \mathcal{O}(\kappa^2E^2).
	\end{equation}
	
	Between the parentheses, we observe the scalar potential $V(z_0, \bar z_0)$, evaluated at the minimum around which we expanded
	\begin{equation}\label{eq:potential_de_base}
		V(z, \bar z) = V_F + V_D = e^\mathcal{K}\left(\frac{\nabla W\bar \nabla \bar W}{\partial\bar\partial\mathcal{K}}-3W\bar W\right) + \frac{\mathcal{P}^2}{2\mathrm{Re}(f)}.
	\end{equation}
	
	In other words, the leading term in $\mathcal{O}(\kappa^2E^4/|m_{3/2}|^2)$ cancels between the three channels when the potential is zero at the minimum. If things are to be consistent, this is the case in our setup, since as mentioned earlier we computed the amplitudes in a Minkowski background. We checked that this result holds for other helicity assignments. In the case where $z$ has a vanishing vev $v=0$, the $U(1)$ is not broken, so the computation involves the complete complex scalar and a massless gauge vector, but the result also holds.
	Therefore, looking at the total amplitude, the perturbative unitarity cutoff is pushed up to the Planck scale $\Lambda\sim1/\kappa\sim M_{\mathrm{Pl}}$. This result can certainly be extended with more than one chiral and vector multiplets, allowing for more general gauge groups, in a Minkowski background.
	It could be interesting to see how it extends to more general backgrounds, such as (anti) de Sitter spacetimes; or even non-static ones, since cosmological setups often involve scalar fields evolving away from the minima of their potential; we leave this for future work.
	
	\vfill\newpage\vfill
	
	\section{Unitarity cutoff with the new Fayet-Iliopoulos terms}
	
	\vfill
	
	Let us now include the new Fayet Iliopoulos terms in our analysis. As mentioned in the introduction, we consider two cases, that we call the \emph{original new FI term} (or FI-I), with non K\"ahler invariant matter couplings \cite{Cribiori:2017laj} and \emph{K\"ahler invariant new FI term} (or FI-II) \cite{Antoniadis:2018oeh, Antoniadis:2019nwz}. They are both based on a vector multiplet, that we denote $V=(A_{\mu},\lambda,D)$ in terms of its components in the Wess-Zumino gauge. We also denote $S_{0}=(s_{0},P_{L}\Omega_{0},F_{0})$ and $\bar{S}_{0}=(\bar{s}_{0},P_{R}\Omega_{0},\bar{F}_{0})$ the chiral and anti-chiral compensator multiplets. In multiplet notations, the Lagrangian of the original new FI term is
	
	\begin{equation}\label{eq:newFIterm}
		\mathcal{L}_{\text{FI-I}}=-\xi_{n}\left[S_{0}\Bar{S}_{0}\frac{w^{2}\bar{w}^{2}}{\bar{T}(w^{2})T(\bar{w}^{2})}(V)_{D}\right]_{D},
	\end{equation}
	
	where $\xi_{n}$ is a parameter and $(V)_{D}$ is the real linear multiplet whose lowest component is $D$, the real auxiliary field of the vector multiplet. The multiplets $w^{2}$ and $\Bar{w}^{2}$ are defined by their lowest component as
	\begin{equation}
		w^{2}=\frac{\Bar{\lambda}P_{L}\lambda}{s^{2}_{0}}\qquad\text{and}\qquad\Bar{w}^{2}=\frac{\Bar{\lambda}P_{R}\lambda}{\bar{s}^{2}_{0}}.
	\end{equation}
	
	Finally the operators $T$ and $\Bar{T}$ are the chiral and anti-chiral projectors (namely, if $(\bar X, P_R\Omega, \bar F)$ is an anti-chiral multiplet, $T(\bar X)$ is the chiral multiplet whose lowest component is $\bar F$). Expanding into components, and in the Poincare gauge where $s_{0}=\Bar{s}_{0}=e^{\mathcal{K}/6}$, we have
	\begin{equation}\label{eq:lagrangian_FI-I}
		e^{-1}\mathcal{L}_\mathrm{FI-I} = -\xi_n e^{{\mathcal{K}}/3} D + \frac{i}{2}\xi_{n}e^{{\mathcal{K}}/3}\Bar{\psi}\cdot\gamma\lambda + \mathcal{O}(\lambda^2).
	\end{equation}
	
	The first term is a contribution to the FI $D$-term; after integrating out the $D$ auxiliary field, it translates into a modification of $V_D$ in \eqref{eq:potential_de_base}, which becomes
	\begin{equation}\label{eq:VDprime}
		V_{D,\text{ FI-I}}=\frac{(\mathcal{P}+\xi_{n}e^{\mathcal{K}/3})^2}{2\mathrm{Re}(f)}.
	\end{equation}
	
	As explained in \cite{Antoniadis:2018cpq}, this new FI term is not K\"ahler invariant. This is clear in \eqref{eq:newFIterm}, because before conformal gauge fixing the compensator $S_0$ transforms as $S_{0}\rightarrow S_{0}e^{J(z)/3}$. Moreover, even though the addition of this FI term does not require R-symmetry, we can consider combining it with the standard FI term discussed after \eqref{eq:lagrangian_N=1}. With the standard FI term, the $U(1)$ is an R-symmetry, so $S_0$ transforms under it, and the new FI term also breaks gauge invariance; except in the particular K\"ahler frame where the $U(1)$ is not an R-symmetry, and where the superpotential is constant, as discussed after \eqref{eq:Kahler_Killing}. In short, if we want to add both the standard and original new FI term at the same time, we should write them in the K\"ahler frame where the superpotential is constant. From there, nothing forbids us to perform K\"ahler transformations, but we shall keep in mind that different K\"ahler frames are not equivalent.
	
	\vfill\newpage\vfill
	
	In order to avoid this issue, a K\"ahler invariant new FI term has been proposed in \cite{Antoniadis:2018oeh}
	\begin{equation}\label{eq:KahlerinvnewFIterm}
		\mathcal{L}_{\text{FI-II}}=-\xi_{n}\left[(S_{0}\bar{S}_{0}e^{-\mathcal{K}/3})^{-3}\frac{(\bar{\lambda}P_{L}\lambda)(\bar{\lambda}P_{R}\lambda)}{\bar{T}(w'^{2})T(\bar{w}'^{2})}(V)_{D}\right]_{D},
	\end{equation}
	
	where the multiplets $w', \bar w'$ are defined by their lowest component as
	\begin{equation}
		w'^{2}=\frac{\Bar{\lambda}P_{L}\lambda}{(s_{0}\bar{s}_{0}e^{-\mathcal{K}/3})^{2}}\qquad\text{and}\qquad\Bar{w}'^{2}=\frac{\Bar{\lambda}P_{R}\lambda}{(s_{0}\bar{s}_{0}e^{-\mathcal{K}/3})^{2}}.
	\end{equation}
	
	This term is now manifestly K\"ahler invariant, because the transformation of $S_{0}\rightarrow S_{0}e^{J(z)/3}$ is accompanied by a transformation of $\mathcal{K}\rightarrow \mathcal{K}+J(z) + \bar J(\bar z)$. As before, we can expand it into components, and get
	\begin{equation}\label{eq:lagrangian_FI-II}
		e^{-1}\mathcal{L}_\mathrm{FI-II} = -\xi_n D + \frac{i}{2}\xi_{n}\Bar{\psi}\cdot\gamma\lambda + \mathcal{O}(\lambda^2)
	\end{equation}
	
	which now leads to
	\begin{equation}\label{eq:D-term_FI-II}
		V_{D, \mathrm{FI-II}}=\frac{\left(\mathcal{P}+\xi_{n}\right)^2}{2\mathrm{Re}(f)}.
	\end{equation}
	
	To this order, the two new FI terms only differ by a factor $e^{\mathcal{K}/3}$ accompanying $\xi_n$. We  therefore introduce the notation $\Delta = e^{\mathcal{K}/3}$ for the original new FI term and $\Delta = 1$ for the K\"ahler invariant new FI term in order to write both cases at once. Actually, in a further generalisation of the  K\"ahler invariant new FI term, the coefficients $\xi_n$ become field dependent functions, invariant under K\"ahler transformations. An explicit example is when their field dependence arises through the K\"ahler invariant combination ${\cal G}={\cal K}+\log|W|^2$~\cite{Antoniadis:2019nwz}. 	Now, when it comes to the $2\rightarrow2$ gravitino scattering amplitudes, it is clear from both  \eqref{eq:lagrangian_FI-I} and \eqref{eq:lagrangian_FI-II} that the new FI terms do not contain gravitino-gravitino-vector nor gravitino-gravitino-scalar interactions, since all the terms in $\mathcal{O}(\lambda^2)$ contain at least two gauginos. Consequently, the inclusion of these terms does not affect the amplitudes of the previous section. One shall just check that it is still possible to use the unitary gauge, to cancel the mixing between the gravitino and the Goldstino. For this, it is enough to check that the Goldstino undergoes a non zero shift under supersymmetry transformations. With the contributions from standard $\mathcal{N}=1$ supergravity and the new FI terms, the Goldstino reads
	\begin{equation}
		P_L\upsilon = -\frac{1}{\sqrt{2}} e^{\mathcal{K}/2}\nabla W\chi - \frac{i}{2}\mathcal{P}P_L\lambda -\frac{i}{2}\xi_{n}\Delta P_L\lambda,
	\end{equation}
	
	and its SUSY variation
	\begin{equation}\label{goldstino_transformation}
		\delta\upsilon_{L}=\frac{1}{2}\left(e^\mathcal{K}\frac{\nabla W\bar \nabla \bar W}{\partial\bar\partial\mathcal{K}} + \frac{\mathcal{P}(\mathcal{P}+\xi_n\Delta)}{2\mathrm{Re}(f)}\right)\epsilon_{L}+\cdots.
	\end{equation}
	
	When $\xi_n = 0$, the term between the parenthesis is always positive in a broken SUSY phase, so we can make a transformation such that $\upsilon_L \rightarrow 0$, which defines the unitary gauge. When $\xi_n\neq 0$, it is no longer the case and one should verify that this parenthesis does not vanish when evaluated at the minimum of the potential around which we compute the amplitudes.
	
	\vfill\newpage\vfill
	
	Let us extend the discussion of section 2 with these new FI terms. As mentioned in the introduction, their presence modify the scalar potential and they don't contribute to the amplitudes, so they generically lead to unanceled leading terms $\mathcal{O}(\kappa^2E^4/|m_{3/2}|^2)$ in the latter and a perturbative cutoff at the SUSY breaking scale. However, this cancellation can be restored at the minimum of the potential for particular values of parameters. The condition for cancellation in the amplitudes is \eqref{eq:divcancel}
	\begin{equation}\label{eq:condition_cancellation}
		V_F(z_0, \bar z_0) + V_{D}(z_0, \bar z_0) = 0,
	\end{equation}
	
	where $V_F$ and $V_D$ are defined in \eqref{eq:potential_de_base}, and $(z_0, \bar z_0)$ is an extremum of the scalar potential
	\begin{equation}\label{eq:condition_extremum}
		\partial(V_F(z_0, \bar z_0)+ V_{D, \mathrm{FI}}(z_0, \bar z_0)) = \bar\partial(V_F(z_0, \bar z_0)+ V_{D, \mathrm{FI}}(z_0, \bar z_0)) = 0,
	\end{equation}
	
	where, with the notation $\Delta$ introduced after \eqref{eq:D-term_FI-II} we can write $V_{D, \mathrm{FI}}$ in the two cases as
	\begin{equation}
		V_{D, \mathrm{FI}}=\frac{\left(\mathcal{P}+\xi_{n}\Delta\right)^2}{2\mathrm{Re}(f)}.
	\end{equation}
	
	In addition, the condition for vanishing cosmological constant at the minimum, as assumed in our amplitude computation, is
	\begin{equation}\label{eq:condition_minkowski}
		V_F(z_0, \bar z_0) + V_{D, \mathrm{FI}}(z_0, \bar z_0) = 0.
	\end{equation}
	
	The conditions \eqref{eq:condition_cancellation} and \eqref{eq:condition_minkowski} straightforwardly imply $2\mathcal{P}(z_0, \bar z_0) + \xi_n\Delta(z_0, \bar z_0) = 0$, in other words that the new FI term does not contribute to the potential at the minimum.
	
	\vfill
	
	It is not hard to find concrete examples where \eqref{eq:condition_cancellation}, \eqref{eq:condition_extremum} and \eqref{eq:condition_minkowski} are simultaneously satisfied. For instance, let us consider the case with K\"ahler potential $\mathcal{K}(z, \bar z) = z\bar z$, superpotential $W(z) = z^b$, gauge kinetic function $f(z) = 1$, a standard FI term $\xi_s$, and either the original or the K\"ahler invariant new FI term with parameter $\xi_n$. The presence of the standard FI term implies that the $U(1)$ is an R-symmetry under which the superpotential transforms with a charge $\xi_s$, so the charge of the scalar satisfies $qb = \xi_s$. Also, as mentioned after equation \eqref{eq:VDprime}, the original new FI term should be written in the K\"ahler frame where the superpotential is constant, with $\tilde{\mathcal{K}}(z, \bar z)= z\bar z + b\log(z\bar z)$. Taking all of this into account, it is not hard to find that for both new FI terms and $0 < b < 0.75$, there is one value of $(\xi_s, \xi_n)$ for which all conditions are satisfied. For $b < 0$, there are two such values. For instance, for the original new FI term
	
	\begin{itemize}
		\item[$\cdot$]$b = 0.5, \quad \xi_s = 0.69211, \quad \xi_n = -2.93863, \quad r_{0} = 0.96407$,
		\item[$\cdot$]$b = -1, \quad \xi_s = 6.43856, \quad \xi_n = -5.04518, \quad r_{0} = 0.62367$,
		\item[$\cdot$]$b = -1, \quad \xi_s = 2.09581, \quad \xi_n = \phantom{-}4.76335, \quad r_{0} = 1.81251$,
	\end{itemize}
	
	where $r_{0} = |z_0|$.
	
	\vfill\newpage\vfill
	
	For the K\"ahler invariant new FI term, we find similarly e.g.
	\begin{itemize}
		\item[$\cdot$]$b = 0.5, \quad \xi_s = 0.50204, \quad \xi_n = -3.92696, \quad r_{0} = 1.20644$,
		\item[$\cdot$]$b = -1, \quad \xi_s = 5.54153, \quad \xi_n = -7.43073, \quad r_{0} = 0.57406$,
		\item[$\cdot$]$b = -1, \quad \xi_s = 1.43121, \quad \xi_n = \phantom{-}8.83171, \quad r_{0} = 2.02123$.
	\end{itemize}
	
	In figure 1 we plot the corresponding potentials for concreteness. In all cases, we checked that the parenthesis in \eqref{goldstino_transformation} does not vanish, allowing us to use the unitary gauge.
	
	\vfill
	
	\begin{figure}[ht!]
		\centering
		\includegraphics[scale=0.5]{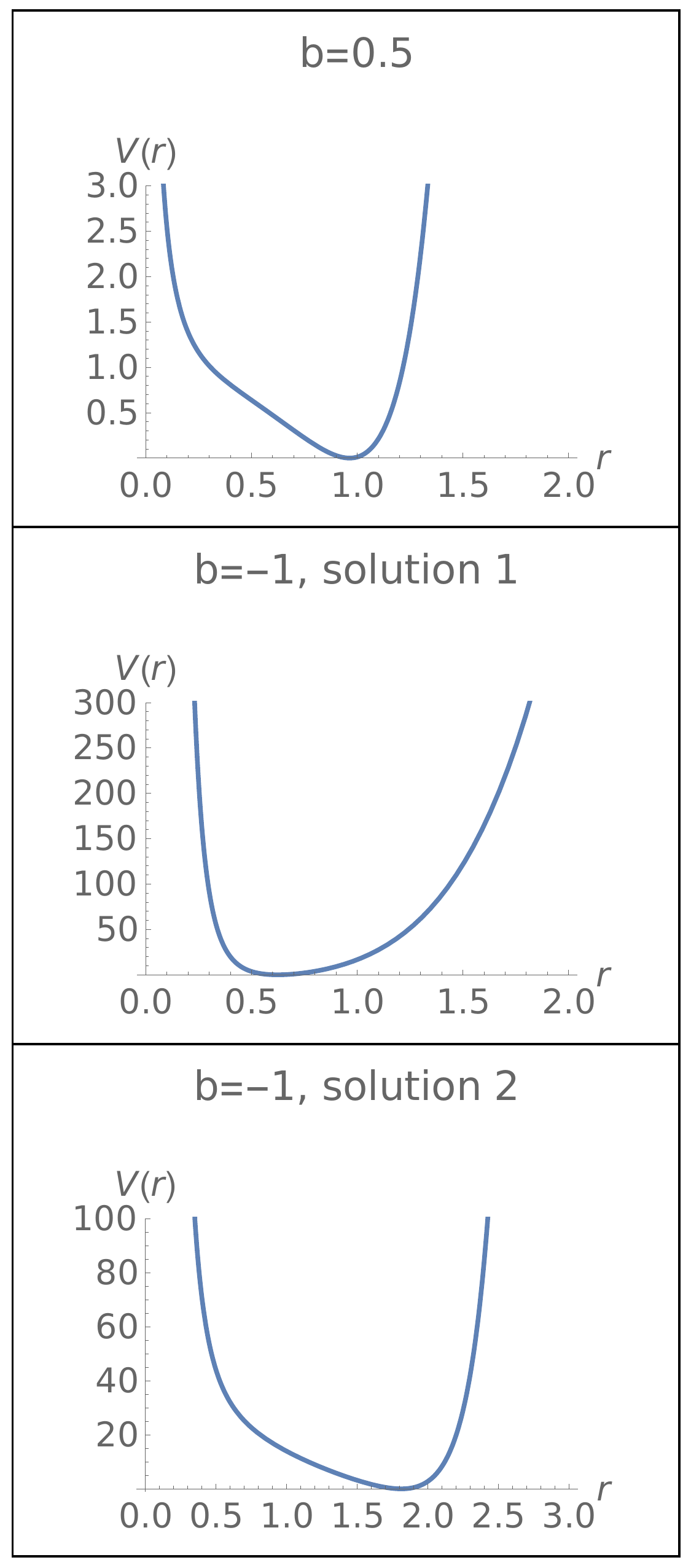}\qquad
		\includegraphics[scale=0.5]{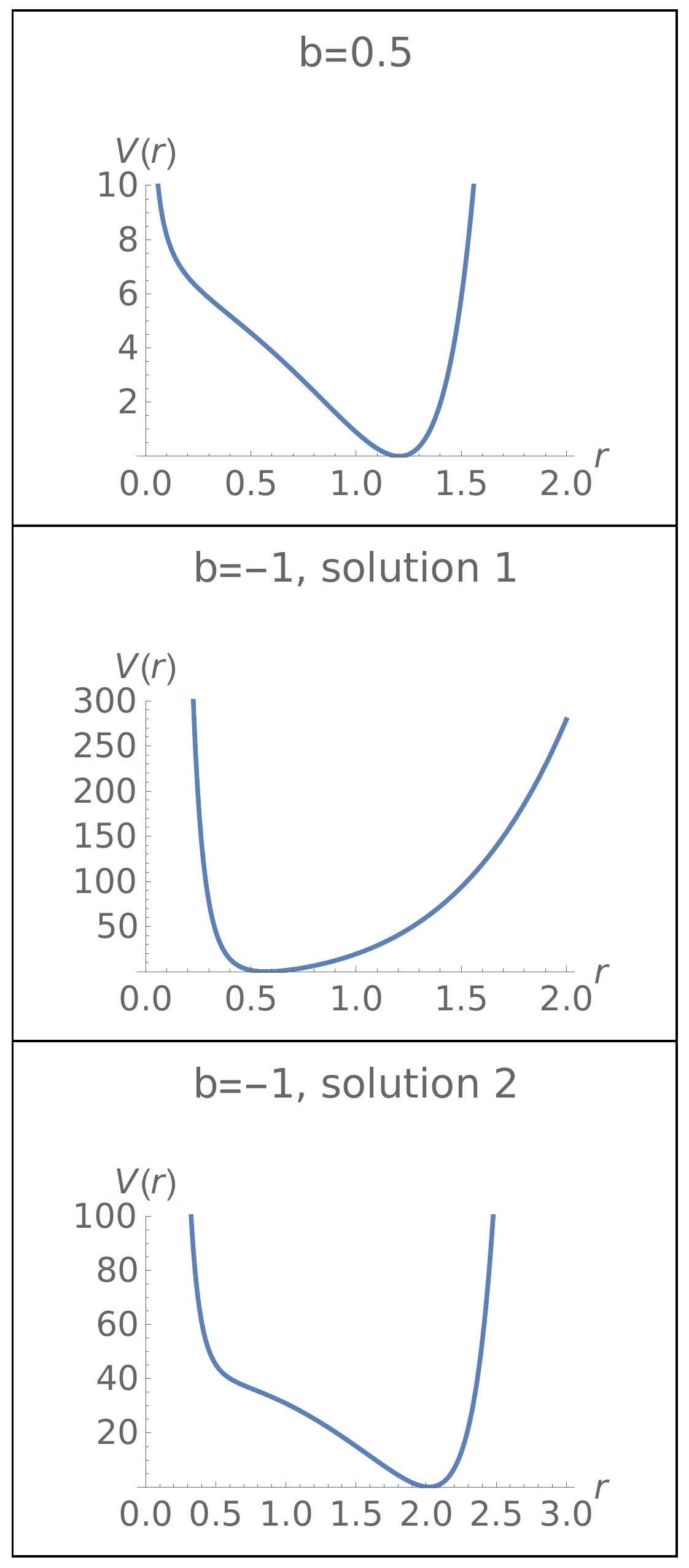}
		\caption{Scalar potential for the model with K\"ahler potential $\mathcal{K}(z, \bar z) = z\bar z$, superpotential $W(z) = z^b$, a standard FI term $\xi_s$, and either the original (left) or the K\"ahler invariant (right) new FI term with parameter $\xi_n$ and values of the parameters for which \eqref{eq:condition_cancellation}, \eqref{eq:condition_extremum} and \eqref{eq:condition_minkowski} are satisfied.}
	\end{figure}
	
	\vfill\newpage\vfill
	
	Of course these particular values of the parameters are not expected to be stable under quantum corrections, unless they are protected by symmetries in some models.
	Except for particular values of the parameters, for which the new FI term does not contribute to the potential at its minimum, as we just saw, unitarity of the $2\rightarrow2$ gravitino scattering amplitude leads to a perturbative cutoff at the SUSY breaking scale with the new FI terms. Note that generalizations of these new FI terms \cite{Antoniadis:2019nwz,Aldabergenov:2018nzd} allow for a field dependent $\xi_n(z, \bar z)$, in which case it is possible to obtain $2\mathcal{P}(z, \bar z) + \xi_n(z, \bar z)\Delta(z, \bar z) = 0$ for all $z$. Then, the new FI term would not contribute to the potential at all, and the discussion of perturbative unitarity would boil down to the standard supergravity case discussed in section 2. However, the phenomenological interest of these new FI terms might be rather limited if they do not contribute to the scalar potential.
	
	%\vfill
	
	\section{Conclusions}
	
	We have shown that the $2\rightarrow2$ massive gravitino scattering amplitudes in ${\cal N}=1$ supergravity has a unitarity breaking cutoff at the Planck scale, even in the presence of the standard FI term associated to a gauged R-symmetry $U(1)$, around a Minkowski minimum of the scalar potential, and when supersymmetry is broken by both $F$- and $D$-term expectation values. This unitarity cutoff is at the Planck scale thanks to cancellation in the amplitudes that happen precisely because the potential vanishes at the Minkowski minimum. We expect this result to hold in more general backgrounds, such as (anti) de Sitter spacetimes, but it could be interesting to check it, since amplitude computations are more subtle in these cases. Non-static backgrounds could also be interesting for cosmological applications, or to study perturbative unitarity during (and not only after) spontanous SUSY breaking. This property of the amplitudes is not valid when a new FI term is added in the action, unless its coefficient is tuned so that the new FI term does not change the value of the potential at its minimum. With this tuning, the new FI term can still contribute to the potential away from the minimum, and it also contains fermionic terms, for instance it contributes to the mass of the physical fermion.
	
	\vfill
	
	\bibliography{bibliographie}
	
\end{document}